\documentclass[twocolumn,prl,amsmath,amssymb,superscriptaddress]{revtex4-1}

\usepackage{graphicx}
\usepackage{dcolumn}
\usepackage{bm}
\usepackage{mathrsfs}
\usepackage{color}
\usepackage{epstopdf}
\usepackage{booktabs}
\usepackage{amsmath}
\usepackage[T1]{fontenc}
\usepackage{calligra}
\usepackage{anyfontsize}
\usepackage[normalem]{ulem}
\usepackage{float}
\usepackage{xcolor}

\usepackage{physics}

\begin{document} 
\title{Topological robustness of orbital angular momentum entanglement in stochastic channels}

\author{Tatjana Kleine}
\affiliation{School of Physics, University of the Witwatersrand, Private Bag 3, Wits 2050, South Africa}

\author{Pedro Ornelas}
\affiliation{School of Physics, University of the Witwatersrand, Private Bag 3, Wits 2050, South Africa}

\author{Cade Peters}
\affiliation{School of Physics, University of the Witwatersrand, Private Bag 3, Wits 2050, South Africa}

\author{Zhenyu Guo}
\affiliation{Centre for Disruptive Photonic Technologies, School of Physical and Mathematical Sciences, Nanyang Technological University, Singapore 637371, Singapore}

\author{Bereneice Sephton}
\affiliation{Dipartimento di Fisica, Universit\`a di Napoli Federico II, Complesso Universitario di Monte S. Angelo, Via Cintia, 80126 Napoli, Italy}

\author{Isaac Nape}
\affiliation{School of Physics, University of the Witwatersrand, Private Bag 3, Wits 2050, South Africa}

\author{Yijie Shen}
\affiliation{Centre for Disruptive Photonic Technologies, School of Physical and Mathematical Sciences, Nanyang Technological University, Singapore 637371, Singapore}
\affiliation{School of Electrical and Electronic Engineering, Nanyang Technological University, Singapore 639798, Singapore}

\author{Andrew Forbes}
\email[email:]{andrew.forbes@wits.ac.za}
\affiliation{School of Physics, University of the Witwatersrand, Private Bag 3, Wits 2050, South Africa}
\email[Corresponding author: ]{andrew.forbes@wits.ac.za}

\date{\today}

\begin{abstract}
\noindent Orbital angular momentum (OAM) entanglement gives access to multiple qubit and high dimensional Hilbert spaces, but is unfortunately susceptible to disturbance, decaying in real-world noisy channels.  Here, we show there is an underlying topology arising from OAM entanglement that is robust to such channels, which we demonstrate using atmospheric turbulence -- exemplary of stochastic or chaotic media. Using a quantum channel with various turbulence strengths, we find the OAM topological observable preserved even though the OAM itself is shown to be highly sensitive to the turbulence. We show this is true for mixed states too, with the OAM topology intact even as the purity of the state decreases due to decoherence.  Our work offers a new perspective on OAM entanglement preservation, and may easily be extended to other spatial bases, degrees of freedom, as well as complex channels, whether static or dynamic.
\end{abstract}
 
\maketitle

\noindent It is well understood that photons can carry intrinsic orbital angular momentum (OAM) through their spatial structure \cite{allen1992orbital}, where a phase of the form $\exp (-i \ell \phi)$ about the azimuth ($\phi$) imbues light with an OAM in the propagation direction of $\ell \hbar$ per photon for any integer $\ell$. A decade later, conservation was exploited for correlation to show that photons can be entangled in their OAM \cite{mair2001entanglement}, for access to multi-dimensional and high-dimensional Hilbert spaces using the spatial mode basis.  Photonic OAM has since been a highly topical form of classical \cite{forbes2021structured} and quantum \cite{forbes2025progress} structured light, fuelling many fundamental studies and applications, and extensively reviewed to date \cite{shen2019optical,wang2021orbital,wang2022orbital,willner2021orbital,franke202230,willner2021perspectives,padgett2017orbital,krenn2017orbital,Erhard2018}.  The promise of a large encoding alphabet and the benefits of dimensionality were arrested by the challenge, identified very early on, that OAM quantum states would not be stable in distorting channels, with atmospheric turbulence a good example due to its random and time varying nature. The decay in OAM entanglement was predicted \cite{paterson2005atmospheric,roux2015entanglement} and experimentally confirmed for both qubits \cite{ibrahim2013orbital} and qudits \cite{zhang2016experimentally}, with deleterious effects on many quantum applications \cite{li2024state,krenn2015twisted,krenn2016twisted}.  This has stimulated several approaches to circumvent the problem, including adaptive optics \cite{sorelli2019entanglement,zhao2020performance}, entanglement distillation \cite{ndagano2019entanglement}, classical correction procedures \cite{ndagano2017characterizing}, the use of partial coherence \cite{phehlukwayo2020influence}, radial carrier envelopes \cite{nape2018self} and mixing degrees of freedom \cite{PhysRevLett.113.060503}, all with limited success.

Here we show that an underlying topology in OAM correlations can be exploited for robustness to stochastic channels.  Although the OAM on which the topology is built is highly unstable, the topological observables remain robust to even strong perturbations.  We use atmospheric turbulence as a topical and extreme example of a chaotic medium and show theoretically and experimentally that OAM topologies remain intact across a variety of states and perturbation strengths.  We extend the results to mixed states too, showing that topology remains intact as the purity of the state decreases. Our work revisits a pressing challenge with a fresh perspective, and easily extended to other quantum channels of similar stochastic nature, i.e., turbid fluids and underwater channels,  revealing a new approach to exploiting OAM entanglement in realistic conditions. 
\begin{figure*}[t!]
    \centering
    \includegraphics[width=\linewidth]{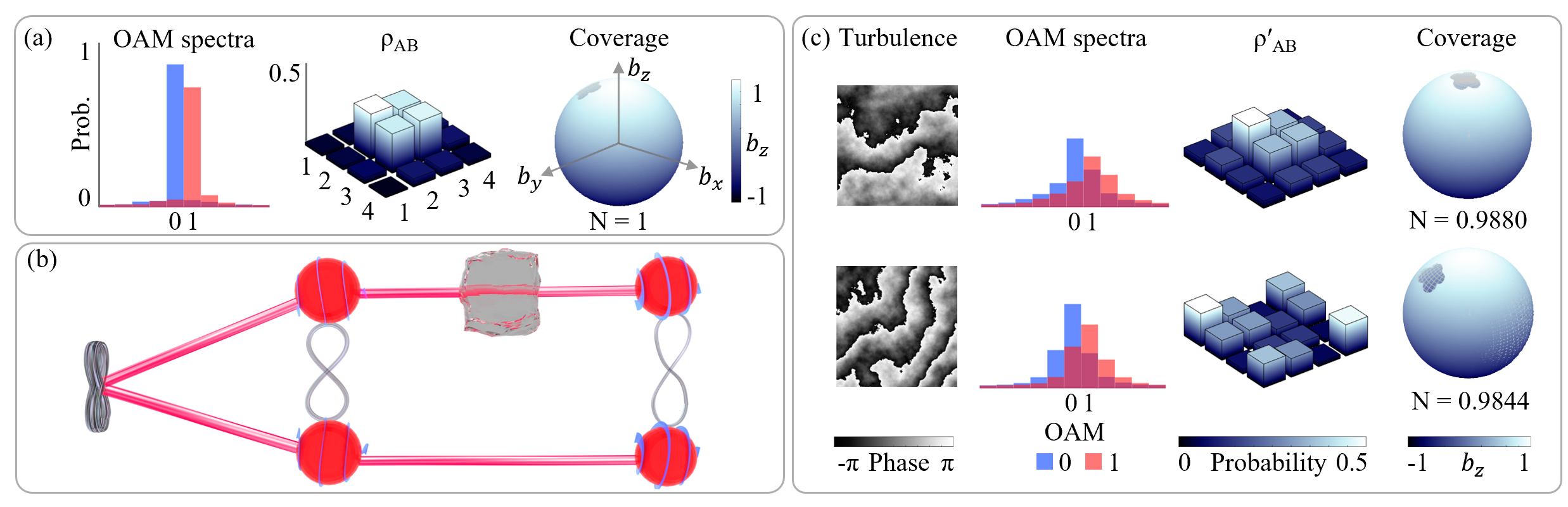}
    \caption{ (a) Reference characterisation of the undisturbed state:  Normalised OAM spectrum showing power in individual OAM modes normalised to the total intensity. Density matrix $\rho_{AB}$ where the label $1$ represents state $|00\rangle$, $2 = |01\rangle$,$3 = |10\rangle$, and $4 = |11\rangle$. Coverage plot constructed from the Bloch vectors $b_{x}$, $b_{y}$ and $b_{z}$ and the calculated topological number. (b) Entangled photon pairs are generated with a defined topology; one photon is transmitted through a turbulent medium, inducing phase perturbations and entanglement decay. (c) Evolution of state properties under varying turbulence strengths. Panels show a representative subset of the ten turbulence screens used. The OAM spectra broaden, the density matrix degrades, local perturbations are seen as a rotation of the coverage sphere, and the topological invariant for each example is shown.}
    \label{Fig:Concept}
\end{figure*}

To begin, consider a bi-photon entangled state of the form $\ket{\psi} = \sum_\ell c_\ell \ket{\ell}\ket{-\ell}$ characterised by Schmidt coefficients $c_\ell$ and expressed in the OAM basis following $\ket{{\pm} \ell} = \int\int |\text{LG}_{0}^{\ell}| e^{i\ell\phi} dr d\phi | \mathbf{r} \rangle$, where $\text{LG}_{0}^{\ell}$ represents a Laguerre Gaussian function with radial index equal to zero ($P=0$). As illustrated by the initial experimental characterisation in Fig.~\ref{Fig:Concept}(a), the undisturbed state has a well-defined OAM spectrum (left panel) where power is concentrated almost exclusively within the targeted modes (OAM states of $\ket{0}$ and $\ket{1}$), and a density matrix (middle panel) that is typical of a pure state. Such states have embedded topologies \cite{koch2025topological}, defined as the map from the real space ($R^2$) of the one photon to the OAM Bloch sphere ($S^2$) of the other. To characterise the topological structure, we project the entangled state onto a position basis to define a unit Bloch vector $\mathbf{b}(\mathbf{r}) = (b_x, b_y, b_z)$ at every spatial coordinate $\mathbf{r}$ of the photon's transverse profile. These vectors are reconstructed via state tomography, where $b_k(\mathbf{r}) = \langle \sigma_{k} \rangle = \langle \psi | ( \sigma_{k} \otimes \ket{\mathbf{r}}\bra{\mathbf{r}} )| \psi \rangle$ represents the expectation value of the $k$-th Pauli operator measured for one photon, conditioned on the spatial position $\mathbf{r}$ of its entangled partner. The surface area covered on the Bloch sphere is described by: ($\epsilon_{ijk}\tilde{b}_i\frac{\partial \tilde{b}_j}{\partial x} \frac{\partial \tilde{b}_k}{\partial y} dx dy$) and is shown in the right panel of Fig.~\ref{Fig:Concept}. The resulting topological number is quantified by integrating the coverage over all real space as follows:
\begin{equation} \label{eq:SkyrmeNumber}
    N = \frac{1}{4\pi}\int_{\mathcal{R}^2} \epsilon_{ijk}\tilde{b}_i\frac{\partial \tilde{b}_j}{\partial x} \frac{\partial \tilde{b}_k}{\partial y} dx dy,
\end{equation}
this describes the integer number of times the Bloch sphere of photon B is wrapped when measuring all states of photon A. The full coverage of the sphere returns a skyrmion number of $N =1$ in this case. 

Now consider the scenario depicted in Fig.~\ref{Fig:Concept}(b) where one of the two entangled photons is passed through some complex channel. The OAM of one photon scatters into other OAM states, resulting in modal cross-talk and entanglement decay, with the dynamics predictable for general complex channels \cite{bachmann2024universal}. Thus, although we will use atmospheric turbulence as our example, we expect the results to be similar in other complex channels. Rather than considering the dynamics of the OAM itself, we now turn our attention to the underlying topology in the OAM entanglement.  Figure~\ref{Fig:Concept}(c), shows the effect turbulence has on the initial state, with two example turbulence cases (rows).  Because of the phase perturbation (left panel), the OAM scatters into adjacent modes and degrades the density matrix, $\rho^{'}_\text{AB}$. The off-diagonal elements, which represent the coherence of the entanglement, diminish as the state structure deviates from its initial pure form. This has a deleterious effect on all traditional OAM quantum observables -- a result well reported in the literature and the crux of the problem to overcome. In contrast, we predict the topology persists, as seen by an inspection of the coverage plot (right panel), where a full coverage of the Bloch sphere represents a non-zero topological charge. Despite local distortions and rotation of the coverage, the vectors continue to span the entire spherical surface, maintaining topological numbers of $0.988$ and $0.984$, respectively, both close to the unperturbed value of $N = 1$. This suggests the topological number is fundamentally decoupled from modal crosstalk induced by the channel and the spreading of the individual photon's OAM, even though it is itself built on OAM to OAM mappings.

To test this experimentally, we subject one photon to a turbulent medium emulated by turbulence screens of varying distortion strength $\Omega$, loaded onto a spatial light modulator (SLM) situated in the near-field of the entanglement source. The strength is defined as $\Omega = 2\omega_{0}/r_{0}$, where $r_{0}$ is the Fried parameter, setting the local statistical fluctuations of the phase screens we employ, and $\omega_{0} = 0.9375$ mm is the back-projected beam waist. 
We investigated eight strengths ranging from $\Omega = 0.25$ to $2.00$, using ten random realisations per strength. We also prepared ten distinct topologies. These included states of the form $|\psi\rangle = \frac{1}{\sqrt{2}}(| 0\rangle_{A}| 0\rangle_{B} + | \ell\rangle_{A}| -\ell\rangle_{B})$ for $\ell \in \{1, 2, 3\}$, as well as a phase-shifted state $|\psi\rangle = \frac{1}{\sqrt{2}}(| 0\rangle_{A}| 0\rangle_{B} + e^{-i\pi/2} | 1\rangle_{A}| -1\rangle_{B})$ and a high-order state $|\psi\rangle = \frac{1}{\sqrt{2}}(| 2\rangle_{A}| -2\rangle_{B} + | 3\rangle_{A}| -3\rangle_{B})$. By swapping the indices of photons A and B, we generated an additional five states with opposite topological charges, yielding $N \in \{\pm 1, \pm 2, \pm 3, \pm 5\}$. Notably, for $|N|=1$, we explored both Neel and saddle-type textures, making it a total of ten states. The spatial structure of these states is experimentally reconstructed by performing local state tomography across the beam profile, where the photons are collected via single-mode fibres (SMFs) acting as spatial filters (see SI for full details).
To establish the statistical limits of this robustness, we measured the average skyrmion number $ N $ across the full range of turbulence strengths using skyrmion numbers measured for ten realisations per strength. As shown in Fig.~\ref{Fig:PureStates}(a), the measured values remain invariant even at the highest distortion levels (see Supplementary for the full dataset of all ten investigated topologies). This experimental stability is faithfully captured by numerical simulations utilising 100 realisations per strength, where the agreement between the experimental markers and the simulated variance (shaded regions) confirms that the topological number is robust against the stochastic medium.

To understand how this invariance is maintained despite local distortions, we examine the evolution of the underlying Bloch field components. In Fig.~\ref{Fig:PureStates}(b), we plot the $b_x$ component of the Bloch field for the $N=3$ state. As the turbulence strength $\Omega$ increases, the field undergoes significant spatial shifting and local deformations. These scrambled features correspond to the modal crosstalk and phase front distortions observed in the OAM spectra. However, because the skyrmion number $N$ is a global property determined by the integral of these field gradients, the local movement of the Bloch vectors does not alter the total winding of the state. This visual evidence demonstrates the clear advantage of topological encoding: while the individual OAM modes are degraded by the channel, the global topological structure is robust to such continuous deformations.

 \begin{figure}[t!]
  \centering
   \includegraphics[width = \linewidth]{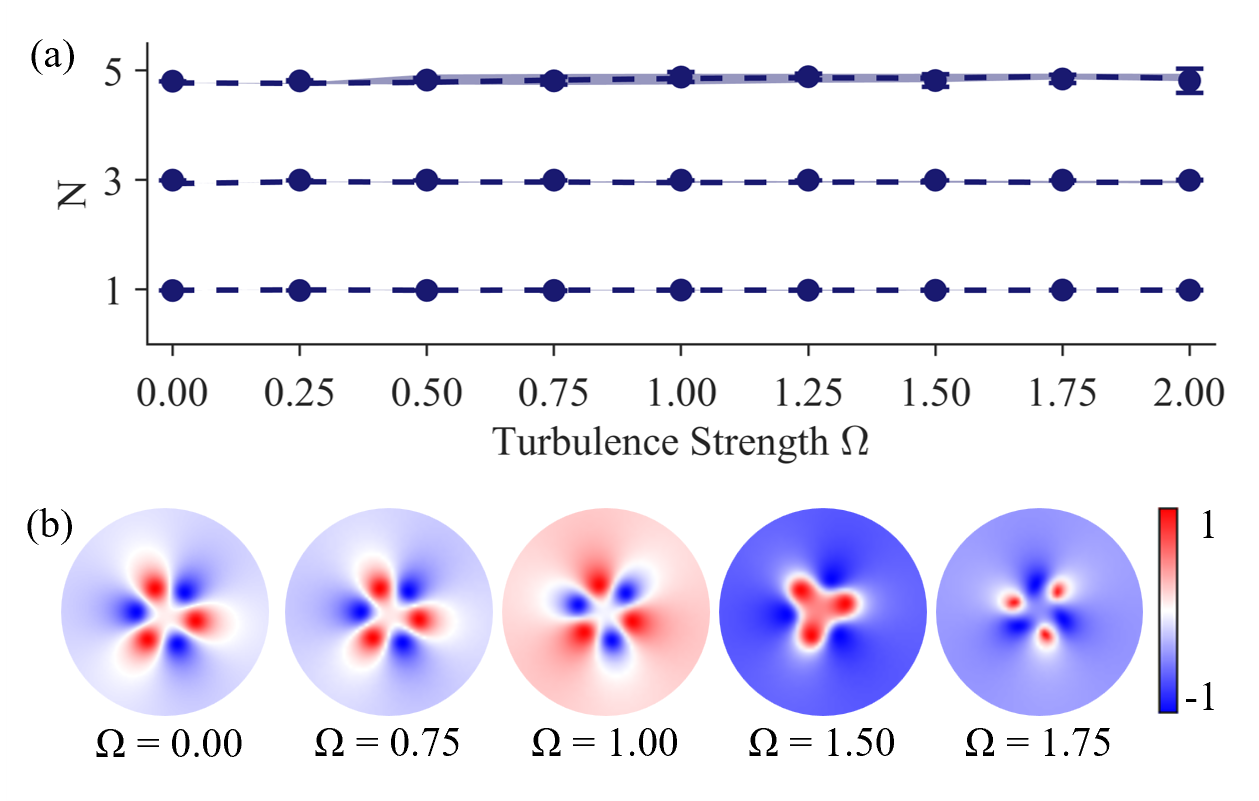}
    \caption{(a) Average measured skyrmion number $ N $ for three distinct topologies as a function of the distortion strength $\Omega$. Experimental data points represent the mean over 10 random turbulence realisations, with error bars indicating the standard deviation. Numerical simulations (dotted lines), averaged over 100 realisations, show excellent agreement with experiment; the shaded regions denote the standard deviation. (b) Reconstructed $b_x$ component of the Bloch field for the state $|\psi\rangle = \frac{1}{\sqrt{2}}(|0,0\rangle + |3,-3\rangle)$ across the investigated range of $\Omega$.} 
   \label{Fig:PureStates}
\end{figure}

\begin{figure*}[t!]
  \centering
   \includegraphics[width = \linewidth]{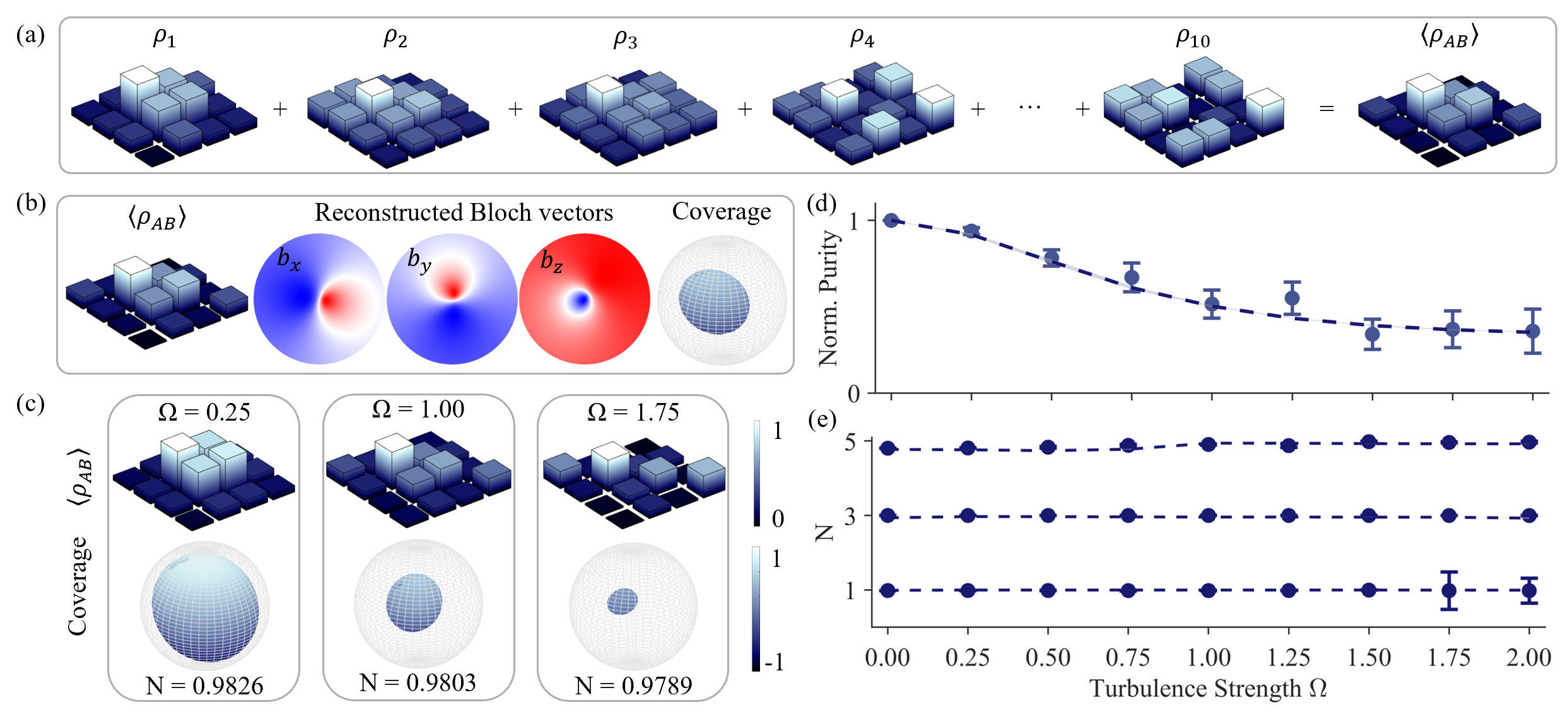}
    \caption{(a)  Schematic of the averaging process, where $n=10$ individual density matrices are summed to compute the ensemble-averaged matrix $\langle \rho \rangle$ for a single turbulence strength $\Omega$. (b) Workflow for the reconstruction of the spatially varying Bloch vector field $\mathbf{b}(\mathbf{r})$ and the skyrmion number $N$ from the mixed state $\langle \rho \rangle$. (c) Experimentally reconstructed density matrices and corresponding coverage plots for the state $|\psi\rangle = \frac{1}{\sqrt{2}}(|0,0\rangle + |1,-1\rangle)$ across the investigated range of turbulence strengths. (d) Normalised state purity $P = \text{Tr}(\langle \rho \rangle^2)$ as a function of turbulence strength, showing a decay toward the mixed-state limit. (e) Measured skyrmion number $N$ extracted from the ensemble-averaged states for various topologies. Markers represent experimental data using 10 realisations, with error bars representing one standard deviation. Dashed lines and shaded regions indicate numerical simulations using 100 realisations and their associated standard deviation.}
   \label{Fig:MixedStates}
\end{figure*}

 Whilst the skyrmionic topology of our states remains invariant for every turbulence realisation, it assumes one can measure the states before they evolve, or that the states do not change with time. However, real-world channels are dynamic and can cause the states to evolve over timescales faster than the measurement integration time. To investigate this scenario, we adopt an ensemble-average approach as depicted in Fig.~\ref{Fig:MixedStates}(a); for each turbulence strength $\Omega$, a single mixed-state density matrix, $\langle \rho \rangle = \frac{1}{n}\sum_{i=1}^{n} \rho_i$ is calculated, by averaging the $n=10$ independent density matrices collected for that strength. In doing so, we model the fluctuating medium as a single stochastic (decoherent) channel.
 
As illustrated by the workflow in Fig.~\ref{Fig:MixedStates}(b) the ensemble-averaged matrix $\langle \rho \rangle$ (left panel), is used to construct the spatially varying Bloch vectors $\mathbf{b}(\mathbf{r})$ (middle panel) and the coverage (right panel).  Although the quantitative skyrmion number is calculated using normalised vectors via Eq.~(\ref{eq:SkyrmeNumber}), unnormalised vectors are used to plot the coverage in  Fig.~\ref{Fig:MixedStates} to better illustrate the decoherent evolution of the state. As turbulence strengths increase while moving from the left to the right panels in  Fig.~\ref{Fig:MixedStates}(c), it can be seen that higher turbulence strengths cause $\langle \rho \rangle$ to evolve toward the identity matrix. This showcases a transition from a pure state toward a completely mixed state. This transition is further visualised through the unnormalized coverage plots in Fig.~\ref{Fig:MixedStates}(c). The magnitude of the local Bloch vectors $| \mathbf{b}(\mathbf{r}) |$ decreases, further illustrating the loss of state purity.

The distinction between the two scenarios is most clearly distinguished by the evolution of state purity $P = \text{Tr}(\rho^2)$  shown in Fig.~\ref{Fig:MixedStates}(d). In the static scenario (Fig.~\ref{Fig:PureStates}), the mean purity of individual realisations remains near unity. In contrast, the ensemble-averaged purity in Fig.~\ref{Fig:MixedStates}(d) decays to $\sim 33\%$, reflecting the transition to a regime dominated by modal crosstalk. 
Remarkably, the topology is preserved, despite this significant loss of purity, further supported by the non-zero quantum discord maintained across all states (see Supplementary). As seen in Fig.~\ref{Fig:MixedStates}(e), the measured skyrmion numbers $N$ remain invariant across all investigated strengths. The experimental markers, which include error bars representing one standard deviation, show agreement with numerical simulations (dashed lines), confirming that the topological number is preserved even amidst a stochastic channel. Fig.~\ref{Fig:MixedStates}(e) confirms that the topological charge is a more resilient observable than the purity of the state itself or the individual OAM correlations.

In conclusion, we have demonstrated that the topological features in OAM-entangled photon pairs show excellent resilience to turbulence, exceeding the stability of the photon's individual OAM states or the state's purity. By transitioning from a static analysis of unitary phase perturbations to an ensemble-averaged framework, we have shown that the skyrmion number remains robust even as the system undergoes decoherence toward the maximally mixed limit. Our work advances the nascent field of quantum optical topologies, moving away from spin-textured fields based on hybrid states \cite{de2025quantum,ornelas2025topological,liu2025nanophotonic,ornelas2024non,koni2025dual} and towards robustness of single degree of freedom entangled states, a new roadmap towards long-distance quantum communication and sensing through fluctuating and chaotic media. By utilising the robustness of the topology, it may become possible to maintain high-dimensional information integrity in stochastic channels (atmospheric, turbid or underwater), providing the stepping stones for topologically-protected quantum protocols in real-world environments. 

\section*{Acknowledgments}
\noindent P.O and C.P acknowledge funding from the Council for Scientific and Industrial Research under the HCD-IBS scholarship scheme.  A.F, I.N and T.K acknowledge funding from the South African Quantum Technology Initiative (SA QuTI). B.S acknowledges the funding from the Italian Ministry of Research (MUR) through the PIRN 2022 project QNoRM and the PNR project PE0000023-NQSTI. Y.S and Z.G acknowledge funding from the Singapore Ministry of Education (MOE) AcRF Tier 1 (RG157/23 \& RT11/23), Singapore Agency for Science, Technology and Research (A*STAR) (M24N7c0080 \& 256I9013), and Nanyang Assistant Professorship Start Up grant

\section*{Data availability}
\noindent The data is not publicly available.
The data is available from the authors upon reasonable
request.

\section{Supplementary Information: Topological robustness of orbital angular momentum entanglement in stochastic channels}

\section{Quantum Experiment}

\subsection{Experimental setup} 
\begin{figure*}[t!]
    \includegraphics[width=0.8\textwidth]{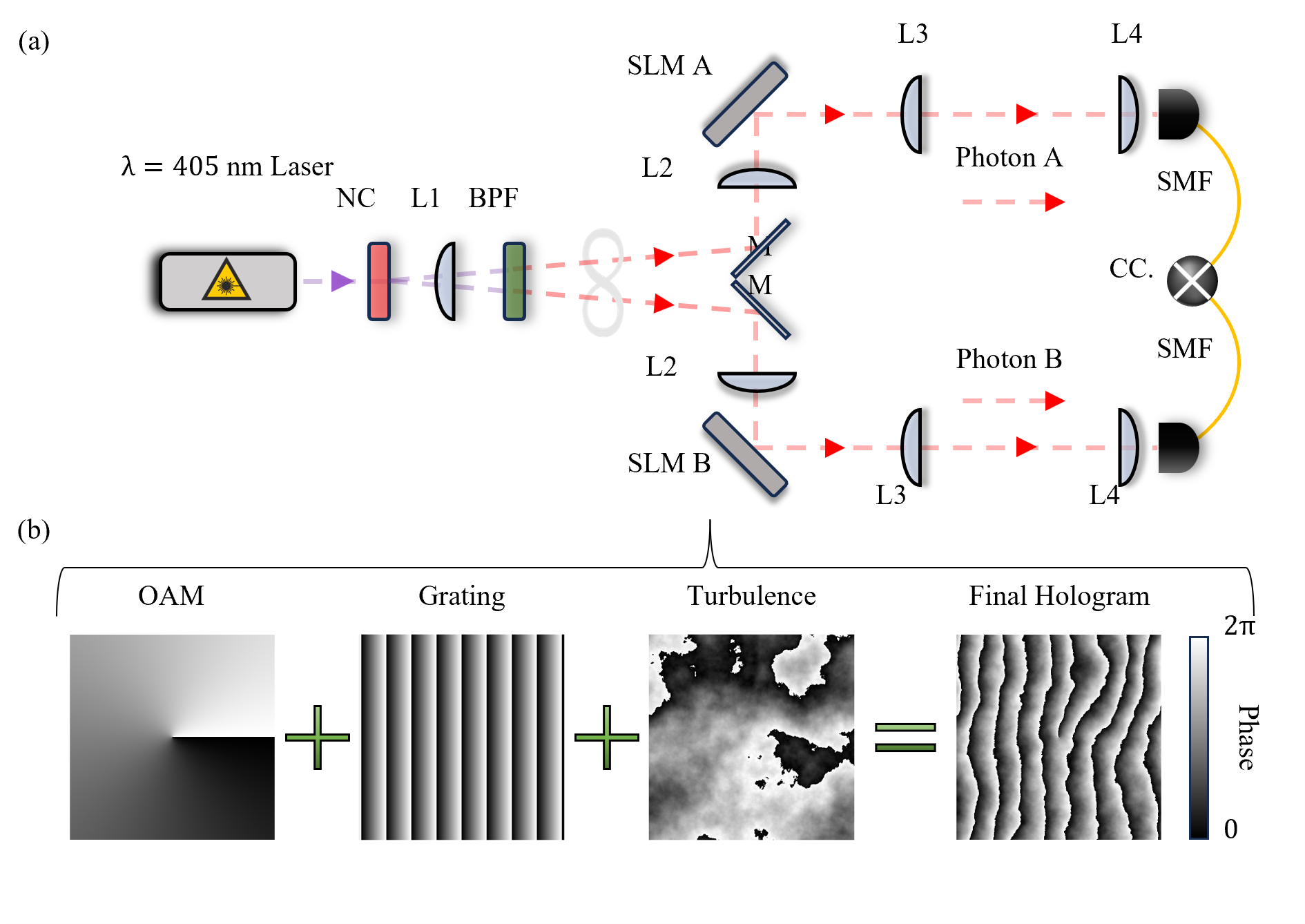}
    \caption{(a) Experimental setup for the generation and detection of OAM-OAM entangled states in turbulence. Abbreviations:  mirror (M), non-linear crystal (NC), lens (L), band-pass filter (BPF), spatial light modulator (SLM), single-mode fibre (SMF), coincidence counter (CC). (b) Construction of phase screens displayed on SLM B, consisting of the spatial projective OAM masks, a grating phase screen and the turbulence phase screen.}
    \label{fig:ExpFig}
\end{figure*}

\noindent A schematic of the experimental setup is shown in Fig.~\ref{fig:ExpFig}. Entangled photon pairs of wavelength $\lambda = 810$ nm were generated through a spontaneous parametric down-conversion (SPDC) process, where an ultraviolet $\lambda = 405$ nm wavelength collimated pump beam  with a Gaussian profile was sent through a type 1, Periodically Poled KTP (PPKTP) non-linear crystal (NC), of length 2 mm. Unconverted pump photons were filtered using a bandpass filter (BPF) with a central wavelength at $810$ nm and a FWHM of $10$ nm. The signal (photon A) and idler (photon B) photons were each imaged and magnified with lenses, L1 and L2, from the crystal plane to the planes of the spatial light modulators, SLM A and SLM B, respectively. The photons were then measured in coincidence by sending them through single mode fibres (SMF) coupled to a pair of single photon detectors connected to a coincidence counter (CC).
The states were characterized by performing a full quantum state tomography (QST), with further details provided in the proceeding section. In this process, spatial projective measurements are performed using a match filter technique comprised of a coupled detection system consisting of the pair of single mode fibres which project each photon along the fundamental Gaussian mode, $|\ell=0\rangle\langle\ell=0|$, and the pair of SLMs. Cycling through different encoded phase holograms on the SLM then effectively rotates the state we wish to measure onto the state accepted by the fibre. Further comprehensive details are provided in the tutorial \cite{nape2023quantum}.\\
 
\noindent Turbulence was introduced in the experiment by encoding phase holograms formed from the combination of an OAM phase hologram, grating hologram, and turbulence phase screen (only on SLM B) as shown in Fig.~\ref{fig:ExpFig} (b). Within a single QST measurement the phase holograms displayed on the SLMs were cycled through different OAM phases whilst leaving the turbulence phase screen and the grating holograms unaltered.

\subsection{Quantum state reconstruction}
\noindent Our biphoton state, $\rho$, is reconstructed by performing a QST in modal space formed by the joint OAM-Hilbert space of both photons, $\mathcal{H}_A\otimes\mathcal{H}_B$. To achieve this, we performed spatially separated projective measurements, $M_{ij}=P^{A}_{i} \otimes P^{B}_{j}$, on each photon, where $P^{A}_{i}$ and $P^{B}_{j}$ are local projections of photon A and photon B on their independent spatial degrees of freedom. The detection probabilities of a system with a corresponding density matrix, $\rho$, are
\begin{equation}
    p_{ij}=\text{Tr}\left(M_{ij}\rho \left( M_{ij} \right)^\dagger\right),
\end{equation}
where $\text{Tr}(\cdot)$ represents the trace operation. In the computational basis, $\{ \ket{0}, \ket{1} \}$, each of the photons are projected onto the basis states $\ket{0}$ and $\ket{1}$ as well as superposition states, $\{ \frac{1}{\sqrt{2}} \left( \ket{0}\pm \exp(i\theta_{A(B)})\ket{1} \right) \}, \theta =  (0,\theta, \pi/2, 3\pi/2)$. In this work the measured 2D OAM states of the form $\ket{\Psi} = \frac{1}{\sqrt{2}}\left(\ket{\ell_1}_A \ket{-\ell_1}_B + \ket{\ell_2}_A \ket{-\ell_2}_B\right)$ where $\ell_{1,2}\geq0$, and we have subsequently mapped them onto the computational basis according to the convention $\{|\ell_1\rangle_A,|\ell_2\rangle_A\}\to\{|0\rangle,|1\rangle\}$ and $\{|-\ell_1\rangle_A,|-\ell_2\rangle_A\}\to\{|1\rangle,|0\rangle\}$. The detection probabilities were then used to determine a guessed matrix, $\rho_G$, which we assumed followed the following decomposition
\begin{equation}
	\rho_G = \frac{1}{4} \big( \mathbb{I}_4 + \sum^{3}_{m,n=1} b_{mn} \sigma_{A,m} \otimes \sigma_{B,n} \big),
\end{equation}
\noindent where $\mathbb{I}_4$ is the four dimensional identity matrix, $b_{mn}$ are the state coefficients and $\sigma_{A,m}$ and $\sigma_{B,n}$ are the Pauli matrices that span the two-dimensional OAM space of both photons, respectively. The density matrix was then calculated using $\rho = \frac{\rho_G^{\dagger}\rho_G}{\text{Tr}(\rho_G^{\dagger}\rho_G)}$ which ensured that it has positive eigenvalues. Here, a least squares fitting procedure was adopted to minimise the measured and predicted detection probabilities, given by
\begin{equation}
    \chi = \sum_{mn} |p_{mn}(\mathbf{b})-p_{mn}^{M}|^2, 
    \label{eq: detProb}
\end{equation}
 where $p_{mn}(\mathbf{b})$ are the detection probabilities for a given density matrix that is determined by the coefficients $b_{mn}$ which are elements of the matrix $\mathbf{b}$ while $p_{mn}^{M}$ are the probabilities that were measured in the experiment.\\ 

 \begin{figure}[t!]
    \includegraphics[width=0.5\textwidth]{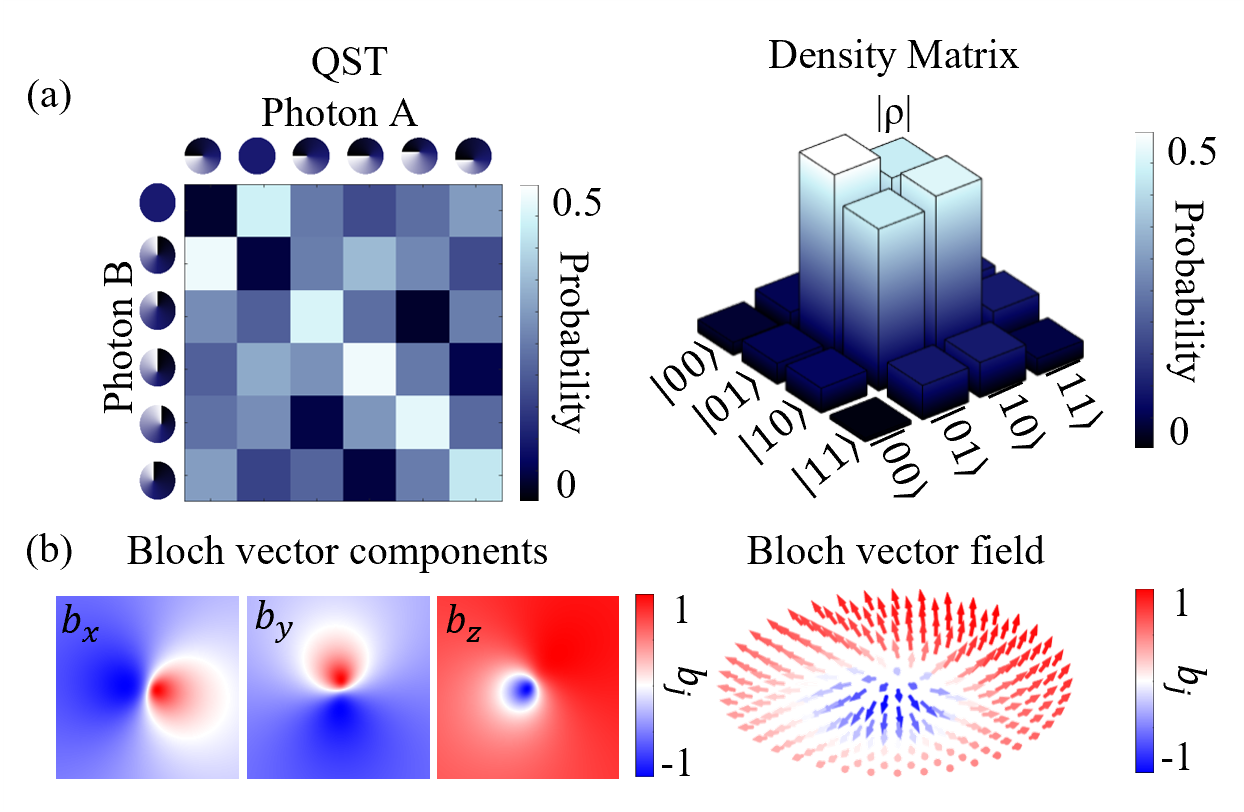}
    \caption{(a) QST (left), associated density matrix (right) and (b) bloch vector components (left) along with the full bloch vector field shown for the state $\ket{\Psi} = \frac{1}{\sqrt{2}}\left(\ket{0}_A \ket{0}_B + \ket{1}_A \ket{-1}_B\right)$.}
    \label{fig:QSTBloch}
\end{figure}
 
An example of a generated QST and associated density matrix is shown in figure \ref{fig:QSTBloch} (a), derived from data captured for the state $\ket{\Psi} = \frac{1}{\sqrt{2}}\left(\ket{0}_A \ket{0}_B + \ket{1}_A \ket{-1}_B\right)$.

\subsection{Statistical analysis}
To quantify the uncertainty in measured Skyrmion numbers, entanglement witnesses and discordance, we quantify the statistical fluctuations in both the single realisation and the ensemble-averaged scenario. In both scenarios, the evolution of the state for each turbulence strength is constructed from $n=10$ independent experimental realisations of said turbulence.\\ 

In the first case, the Skyrmion numbers and entanglement witnesses are taken as an average over the 10 reconstructed density matrices; the reported errors for these quantities are taken from the standard deviation over all realisations. In this scenario, we assumed a detection system capable of measuring within a single realisation of turbulence; therefore, the Skyrmion numbers and entanglement witnesses can be discerned for each individual realisation of turbulence.
However, in the case of the statistical ensemble, we assume the turbulence varies in time faster than we can measure. In such a scenario, we measure a statistical mixture of various states. Therefore, to determine the statistical fluctuations of our measured entanglement witnesses, Skyrmion numbers and discordance, we quantified the fluctuation in the elements of the density matrix, $\rho_{ij}$, with time, i.e., over all realisations. This allows us to quantify the upper and lower extreme bounds that the elements of $\rho$ fluctuate between for any given turbulence strength. Then, by plugging in these boundary values for every calculation, we can extract the upper and lower limits that each quantity oscillates between.\\

\section{Entanglement Witnesses}

\noindent Concurrence, fidelity and purity serve as typical entanglement witnesses quantifying different aspects of the quality of our experimentally measured states. In this work the effect of turbulence is quantified through how these entanglement witnesses decay in the presence of growing turbulence strength. \\

\noindent The concurrence was used to measure the degree of entanglement between the OAM-entangled photons. It was measured from
\begin{equation}
    C(\rho) = \text{max} \{ 0, \lambda_1 -\lambda_2- \lambda_3 - \lambda_4 \},
\end{equation}
where $\lambda_i$ are eigenvalues of the operator $R = \text{Tr} \left( \sqrt{  \sqrt{\rho} \tilde{\rho} \sqrt{\rho}  }  \right)$ in descending order and $\tilde{\rho} = \sigma_{y} \otimes \sigma_{y} \rho^* \sigma_{y} \otimes \sigma_{y}$ with $\sigma_y = \begin{bmatrix}
  0 & -i\\ i & 0
\end{bmatrix}.$ The concurrence ranges from 0 for separable states to 1 for maximally entangled states.\\

\noindent The fidelity is a measure of the distance between any two density matrices. In this work we are interested in quantifying the change in fidelity between the experimentally measured state under ideal experimental conditions, i.e. no turbulence, ($\rho_T$) and the state after passing through a turbulent channel ($\rho$). The fidelity is then defined as
\begin{equation}
    F =\left( \text{Tr}  \left( \sqrt{  \sqrt{\rho_T}\rho \sqrt{\rho_T}  }  \right) \right)^2.
\end{equation}
 The fidelity is then 1 if the states are identical up to a global phase.\\ 

\noindent Lastly, the purity of the state is quantified as
\begin{equation}
    \gamma = \text{Tr}(\rho^2).
\end{equation}
where $\gamma=1$ for a pure state and $\gamma=\frac{1}{4}$ for a mixed state. For static turbulence the transformation can act as a non-unitary transformation within the truncated Hilbert space $\mathcal{H}_A\otimes \mathcal{H}_B$. Therefore whilst we expect the concurrence to degrade within this truncated Hilbert space, the purity of the state should not change. However, in the case of dynamic turbulence, the state evolves as an ensemble average over many density matrices, which means that the coherence between each basis mode is diminished. Quantitatively, this results in the suppression of off-diagonal terms in the density matrix thereby diminishing the purity of the state and driving it toward the identity matrix, which corresponds to a maximally mixed state.

\subsection{Discordance}
We quantified the quantum correlations that persist between our two photons using quantum discord, a measure of the extent to which the system exhibits correlations beyond those achievable in a classical system. The discordance was measured for each state investigated at all turbulence strengths. To achieve this, the quantum correlations in the reconstructed density matrix of our bipartite system, $\rho_{AB}$, were obtained by measuring the difference between the total correlation (Quantum mutual information) and the maximum classical correlation extractable via the local measurements on photon B. The total correlation $I(\rho_{AB})$ was computed using the von Neumann entropy $S(\rho) = -Tr(\rho \log_{2} \rho)$:
\begin{equation}
    I(A:B) = S(\rho_A) + S(\rho_B) - S(\rho_{AB}),
\end{equation}
where $\rho_{A}$ and $\rho_{B}$ are the reduced density matrices obtained via the partial trace: $\rho_{A}= Tr_{B}(\rho_{AB})$ and $\rho_{B}= Tr_{A}(\rho_{AB})$.
The classical correlation $J(A|B)$ is defined as the maximum information gained about photon A by performing a complete set of local projective measurements $\{\Pi_k^B\}$ on photon B:
\begin{equation}
    J(A|B) = \max_{\{\Pi_k^B\}} \left[ S(\rho_A) - \sum_k p_k S(\rho_{A|k}) \right].
\end{equation}
Here, $p_k = \text{Tr}[(\mathbb{I}_A \otimes \Pi_k^B) \rho_{AB}]$ is the probability of the $k$-th outcome, and $\rho_{A|k} = \frac{1}{p_k} \text{Tr}_B[(\mathbb{I}_A \otimes \Pi_k^B) \rho_{AB} (\mathbb{I}_A \otimes \Pi_k^B)]$ is the post-measurement state of $A$.
\begin{figure*}[t!]
    \includegraphics[width=0.9\textwidth]{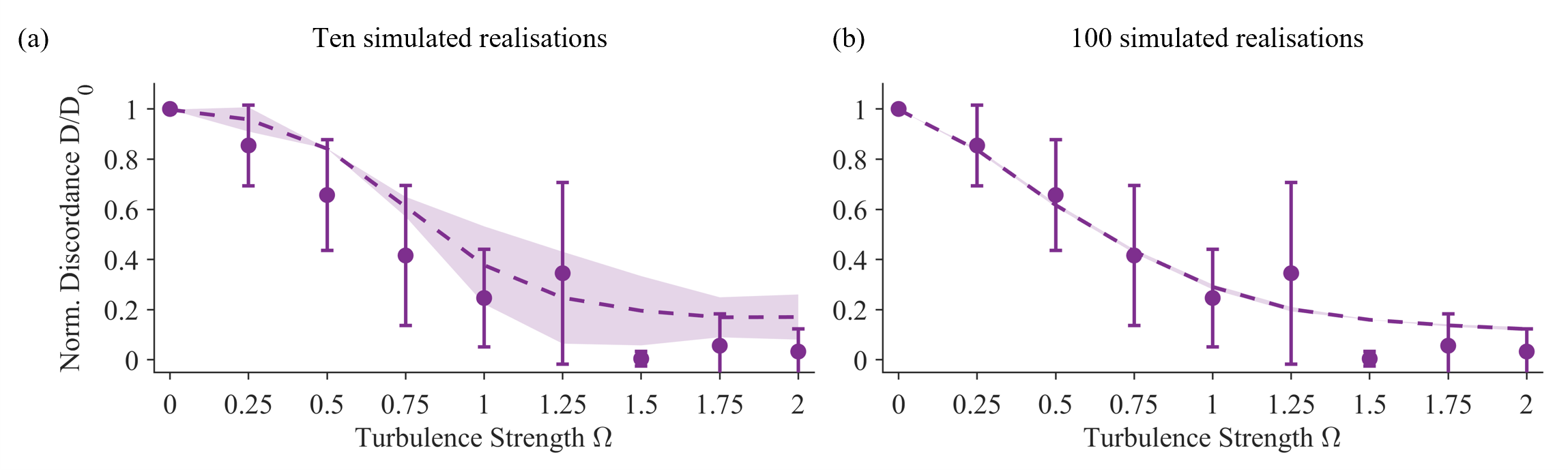}
    \caption{(a). Quantum discordance for the state $|\psi\rangle = \frac{1}{\sqrt{2}}(| 0\rangle_{A}| 0\rangle_{B} + |1\rangle_{A}|-1\rangle_{B})$. Markers represent the mean of 10 experimental realisations, with error bars indicating the standard deviation. Dotted lines denote numerical simulations averaged over 10 realisations. (b). Quantum discordance for the same state. Markers represent the mean of 10 experimental realisations, with error bars indicating the standard deviation. Dotted lines denote numerical simulations averaged over 100 realisations.}
    \label{fig:Discordance}
\end{figure*}
For a two-qubit system, the measurement operators $\Pi_k^B$ are parametrised on the Bloch sphere by the angles $\theta \in [0, \pi]$ and $\phi \in [0, 2\pi]$:
\begin{equation}
    |\psi(\theta,\phi)\rangle = \cos(\frac{\theta}{2})|0\rangle + e^{i\phi}\sin(\frac{\theta}{2})|1\rangle.
\end{equation}
The optimisation was performed using a sequential quadratic programming (SQP) algorithm. A multi-start strategy was used to ensure the algorithm identified the global maximum of $J(A|B)$ rather than a local maximum. The optimiser was initialised from eight distinct points across the $(\theta,\phi)$ parameter space. Finally, the quantum discord was extracted as follows:
\begin{equation}
    D(A|B) = I(A:B)- \underset{\theta,\phi}{\max}J(A|B).
\end{equation}
Quantum discord was extracted from all experimental data and simulated data. For the experimental results, ten realisations were used, resulting in large error bars that represent one standard deviation. This was replicated in simulations using ten realisations as shown in Fig.~\ref{fig:Discordance} (a).

Once the number of realisations used in the simulation was increased to 100, the standard deviation decreased (represented by the shaded region around the dotted line), and the simulated data (dashed line) matched the experimental data found using the ten realisations (dots with error bars). However, the error bars on our experimental data do not change, as only ten realisations were used. Both plots show the normalised discordance; normalisation was included to illustrate the trends better. To normalise the data, the discordance found for a state subject to no turbulence was taken to be $D_{0}$, and all other subsequent discordance values were divided by this value. The error bars were scaled accordingly. Crucially, the discordance for the lowest experimental points was approximately $0.03$ whilst for the simulation it was $0.13$. For all other investigated subspaces, both the experimental and simulated discordance at the strongest turbulence strength ($\Omega = 2.00$) are non-zero, indicating that quantum information remains present at high turbulence strengths. 

\subsection{Quantum Contrast}
\noindent At higher turbulence strengths, we expect our signal to diminish as the probability of measuring a photon in any given OAM state spreads into higher-order subspaces that do not form part of our 2D subspaces of interest. To ensure that our measurement is not dominated by noise, we compute the quantum contrast (QC), which serves as a measure of the signal-to-noise ratio of our system by comparing the number of true coincidental events ($C$) to accidental events ($A$). In general, the quantum contrast for any measurement is quantified using

\begin{equation}
    QC = \frac{C}{A} = \frac{C}{\tau  S_{A}  S_{B}},
\end{equation}

$\tau=2 \text{ns}$ is the gating time, $(S_{A})$ is the total number of counts at detector A, and $(S_{B})$ is the total number of counts at detector B. We note that the reported QC for each realisation of turbulence is the mean QC, $\overline{\text{QC}}$, calculated across all $36$ entries of the QST measurement.
 
 Additionally, to account for the stochastic nature of the turbulence, the final reported QC ($\text{QC}_{\text{avg}}$) and its associated error, ($\text{QC}_{\text{err}}$) are computed as the mean and standard deviation across $n=10$ independent realizations for each turbulence strength $\Omega$:
\begin{equation}
    \text{QC}_{\text{avg}}(\Omega) = \frac{1}{N} \sum_{n=0}^{N-1} \overline{\text{QC}}_n.
\end{equation}
The quantum contrast and errors for each turbulence strength and measured subspace are plotted below in Fig.~\ref{fig:QC_Iter}. From this we note that while there is an observed drop in QC with increasing $\Omega$, the QC never drops below $\text{QC}=10$, which is much greater than the lower limit of $1.5$ at which point the signal is indistinguishable from noise. Therefore, we can assert that our signal dominates over noise for the entire turbulence strength regime under investigation.
\begin{figure}[t!]
    \includegraphics[width=0.5\textwidth]{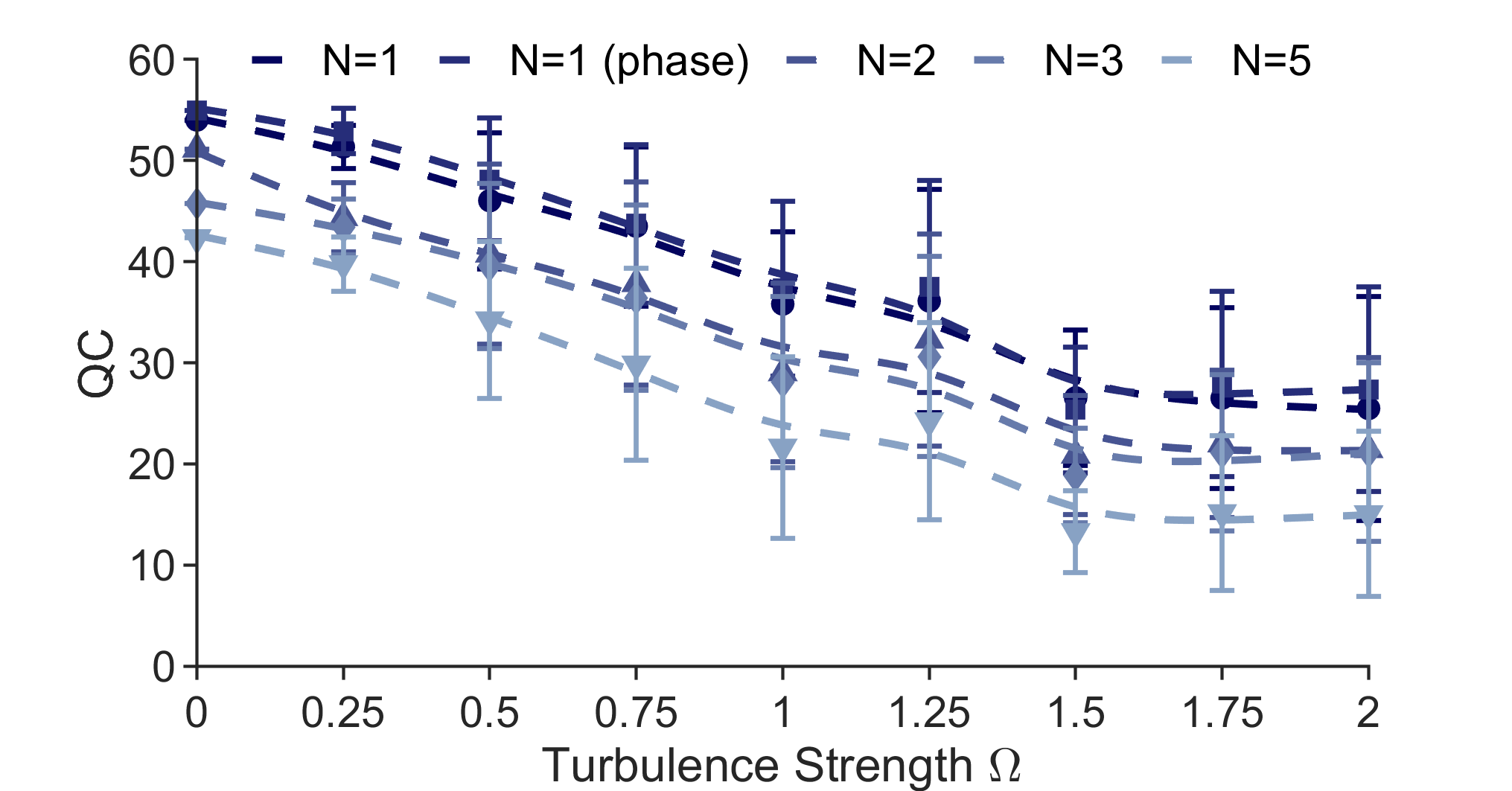}
    \caption{Loss of signal-to-noise ratio quantified through the drop in quantum contrast with increasing turbulence strength for 5 different states denoted with different marker styles and colours as shown in the legend at the top of the plot. Dotted lines indicate the general trend for each subspace, while points with error bars report the QC and error for each turbulence strength.}
    \label{fig:QC_Iter}
\end{figure}

\subsection{Skyrmion number calculation}
\begin{figure}
    \centering
    \includegraphics[width=1\linewidth]{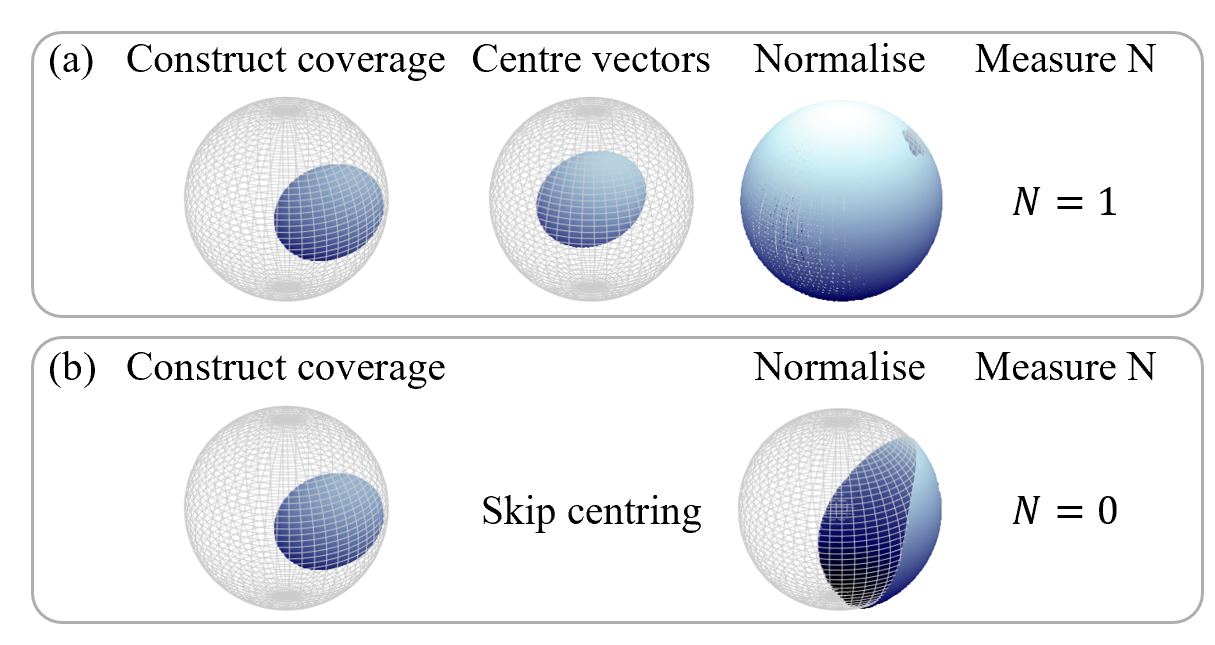}
    \caption{(a). Post-processing procedure showing how to handle Bloch vectors that shift and rotate. The vectors are centred and then normalised before being used to extract the skyrmion number. (b). Overview of the process if the centring step is skipped. The normalisation process can collapse the spatially varying Bloch vectors, thus resulting in a non-integer or zero for the Skyrme number being recovered.}
    \label{fig:No centring}
\end{figure}

\noindent To extract the topological observables from the reconstructed density matrix, we first need to express one of the photons in the position basis. Since OAM-OAM entangled photons are the subject of this work, the choice of the photon which must undergo a basis change is arbitrary, with the main consequence being a change in the sign of the Skyrmion number. This operation can be seen as follows $\rho \to \rho(\vec{r}) = {}_A\langle r |\rho | r \rangle_A$, where $$ \rho = \sum_{ijmn}c_{ijmn}\ket{\ell_i}_A\ket{\ell_j}_B{}_A\bra{\ell_m}{}_B\bra{\ell_n}. $$ Then using $\langle r|\ell_i\rangle = \text{LG}_{\ell_i}(\vec{r})$, the spatially varying density matrix becomes $$ \rho(\vec{r}_A) = \sum_{im}\left[\text{LG}_{\ell_i}(\vec{r}) \, \text{LG}^*_{\ell_m}(\vec{r}) \right] \sum_{jn} c_{ijmn} \ket{\ell_j}_B{}_B\bra{\ell_n}, $$ where the superpositions of complex scalar fields, $\text{LG}_{\ell_i}(\vec{r}) \, \text{LG}^*_{\ell_m}$, become the field coefficients of the density matrix whose elements are labelled $\ket{\ell_j}_B{}_B\bra{\ell_n}$. From the spatially varying description of the density matrix, we can then extract the spatially-varying bloch vector components using $b_i(\vec{r}) = \text{Tr}\left(\sigma_i\rho(\vec{r}) \right)$, where $\sigma_i$ are the typical Pauli spin matrices. Finally, the Skyrmion number is obtained directly from the expression

\begin{equation}
    n = \frac{1}{4\pi}\int_{\mathcal{R}^2} \epsilon_{ijk}\tilde{b}_i\frac{\partial \tilde{b}_j}{\partial x} \frac{\partial \tilde{b}_k}{\partial y} dx dy,
    \label{eq:Skyrme Number}
\end{equation}

where $\tilde{\vec{b}}\cdot \tilde{\vec{b}} = \frac{\vec{b}}{|\vec{b}|} = 1$. We note that some care should be taken before normalising the bloch vectors of the state as illustrated in Fig.~\ref{fig:No centring}.
Noise is capable of not only shrinking and rotating the image of the map $\vec{b}$ as shown in the left panel of Fig.~\ref{fig:No centring} (a) but is also capable of shifting it within the bloch sphere such that the origin point $\vec{b}_0 = (0, 0, 0)^T$ is no longer contained within the geometric object traced out by $\vec{b}$. In such a case, the map should be shifted back towards the origin as shown in the centre panel prior to normalising the $\vec{b}$. Should the map not be shifted back to the origin, normalisation can collapse the vectors as shown in the left panel of Fig.~\ref{fig:No centring} (b) and lead to erroneous errors. A more detailed discussion has been provided in ref. \cite{de2025quantum}.

\section{Emulating turbulence with digital holograms}

\subsection{Modelling atmospheric turbulence}
\label{Section: ModAtmos}

\begin{figure*}[t!]
    \includegraphics[width=0.8\textwidth]{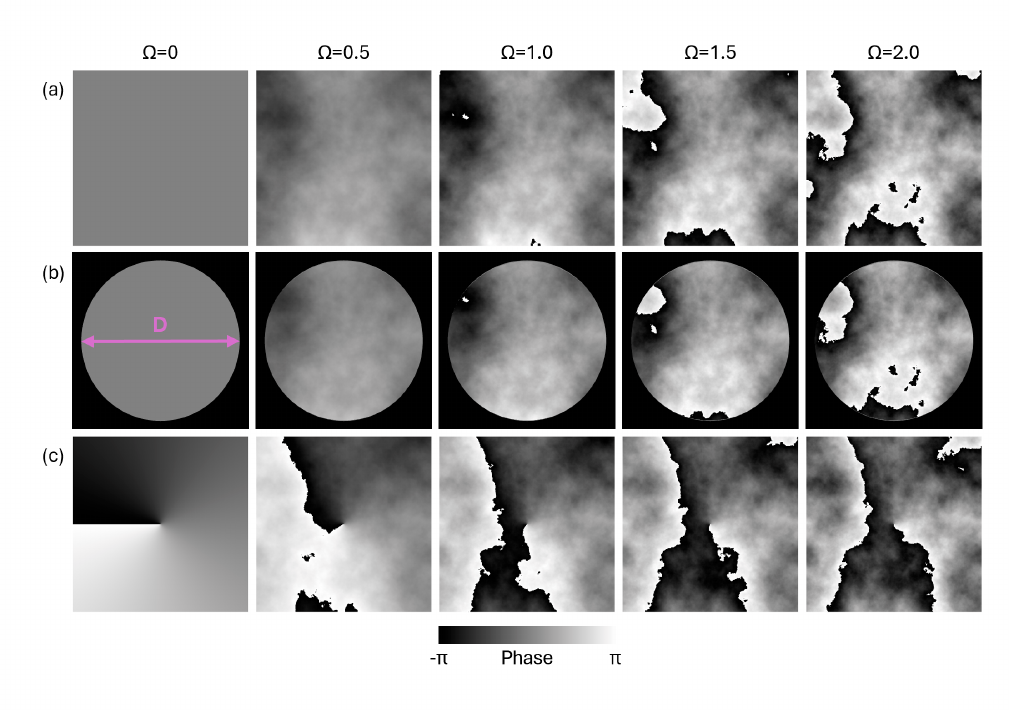}
    \caption{(a) Example turbulence phase screens of various strengths ranging from $\Omega = 0$ to $\Omega = 2.0$. (b) Turbulence phase screens of various strengths overlayed with the beam size $D$. (c) OAM phase profile with $\ell = 1$ after being exposed to turbulence phase screens of various strengths.}
    \label{fig:TurbScreens}
\end{figure*}

The effects of atmospheric turbulence were emulated with the use of digital holograms encoded on a liquid crystal spatial light modulator (SLM). The device allows for pixel-level control of the phase imparted onto an incident complex light field. This enabled us to closely replicate the refractive index distortions that a photon would experience when travelling through a real-world turbulent channel with the use of a phase screen encoded onto the SLM. The phase screens were generated according to the method described in Ref. \cite{peters2025structured}, where full mathematical details and MATLAB code for the method can be found. We made use of the Fourier phase screen method due to its computational efficiency, which randomly samples coefficients in frequency space and weights them according to a model-specific power spectral density (PSD) function. The use of an inverse Fourier transform of these coefficients generates the phase screen in position space, which can then easily be encoded onto the SLM. The Fourier phase screens leveraged subharmonic sampling to better emulate the long-range correlations observed in atmospheric turbulence, and were calculated up to 5 subharmonics.

We opted to use the Kolmogorov PSD, $\Phi(k)$, due to its extensive study in the literature and its well-known analytical relations with structured modes. It is given by,
\begin{equation}
   \Phi(k) = 0.023 r_0^{-5/3}k^{-11/3}\,,
   \label{eq:kol PSD}
\end{equation}
where $k$ represent the magnitude of the angular frequency coordinate $k=\sqrt{k_x^2+k_y^2}$ and $r_0$ is the Fried parameter which represents the average correlation length of phase distortion. We show example turbulence phase screens in Fig.~\ref{fig:TurbScreens} (a). Stronger turbulence strengths are characterised by a smaller Fried parameter as the phase distortions will be correlated over shorter distances. Weaker turbulence is characterised by a large Fried parameter as the phase distortions will be correlated over longer distances. To completely quantify the turbulence strength of a channel, one must also account for the beam size, as it is the size of the complex field relative to the medium's correlation length that determines the severity of the induced distortion. We therefore quantify the strength of the turbulence we encode using the unitless parameter $\Omega = 2w/r_0$, which gives the ratio of the mode diameter $2w$ to the Fried parameter, where $w = w_0 \sqrt{|\ell| + 1}$. We illustrate this visually in Fig. \ref{fig:TurbScreens} (b). Values where $\Omega <1$ indicate weak turbulence, where the beam size is smaller than $r_0$, resulting in minor changes to phase of the incident mode. Values where $\Omega >1$ indicate strong turbulence, where the beam size is larger than $r_0$, resulting in significant changes to the phase of the incident mode. Values where $\Omega \approx1$ can be considered moderate in strength. We show the effects of these phase screens onto an OAM phase profile in Fig.~\ref{fig:TurbScreens} (c).

\subsection{Calibration of experimental turbulence strengths}

\begin{figure*}[t!]
    \includegraphics[width=0.9\textwidth]{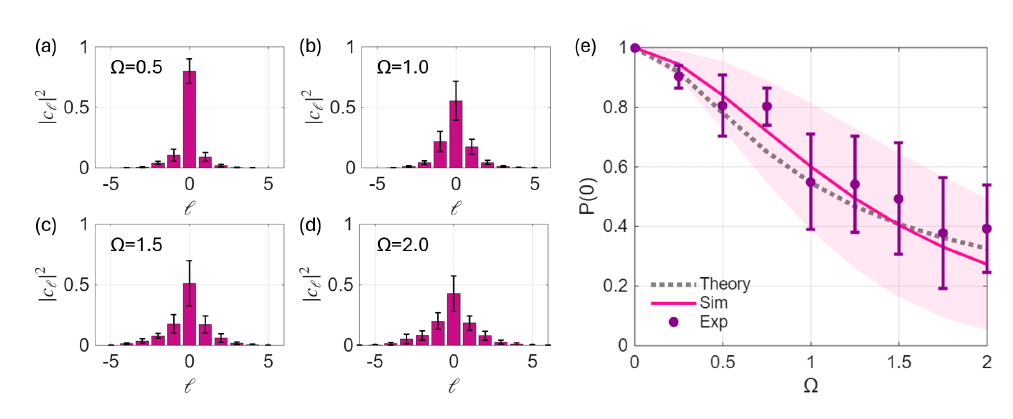}
    \caption{The experimentally measured OAM spectrum of a photon, initially with $\ell = 0$, after experiencing turbulence of strengths (a) $\Omega = 0.5$, (b) $\Omega = 1.0$, (c) $\Omega = 1.5$ and (d) $\Omega = 2.0$. The spectra are averaged over 10 independent realisations of turbulence and the error bars indicate the standard deviation. (e) The probability of measuring the photon with an OAM of $\ell = 0$ as a function of turbulence strength. The simulations are averaged over 100 independent realisations of turbulence with the shaded region indicating the standard deviation. The experimental points are average over 10 independent realizations with the error bars indicating the standard deviation.}
    \label{fig:Calibration}
\end{figure*}

To ensure the encoded phase screens on the SLM imparted distortions of the desired strength, we make use of an analytical result directly related to the measured OAM spectrum of the photon after passing through the turbulent channel. In Kolmogorov turbulence, the probability of a photon that initially had zero OAM being measured to still have zero OAM after propagating through a turbulent channel $P(0)$ is given by Equation 16 in Ref. \cite{klug2021orbital},
\begin{equation}
    P(0) \approx [I_0(\beta) + I_1(\beta)]e^{-\beta},
    \label{eq:calibration}
\end{equation}
where $\beta = 1.8025\Omega^{5/3}$ and $I_n(\cdot)$ is the modified Bessel function of the first kind. We show the experimental results of this calibration in Fig.~\ref{fig:Calibration}. Fig.~\ref{fig:Calibration} (a)-(d) shows the experimentally measured OAM spectra for a photon, initially with $\ell = 0$, after passing through atmospheric turbulence of strengths $\Omega = 0.5,\,1.0,\,1.5$ and $2.0$. These were obtained by encoding the turbulence screen on one arm (SLM B) and then projecting into the various OAM basis states $\ell \in [-10, 10]$ on the other arm (SLM A). The plots show $\ell \in [-5, 5]$ for clarity, as contributions outside that range were negligible. The plots show the average measured OAM spectra over 10 independent realisations of turbulence, with the error bars indicating the standard deviation. As expected, the spectrum broadens around $\ell = 0$ and $P(0)$ decreases as the turbulence strengths increases; a result of the increased probability of imparting non-zero OAM onto the incident photon. Fig.~\ref{fig:Calibration} (e) shows $P(0)$ as a function of $\Omega$. The theoretical prediction, according to Equation~\ref{eq:calibration} is plotted as a dotted gray line. Simulated results are plotted as a solid pink line. The simulations results are average over 100 independent phase screen realisations at each datapoint, with the shaded regions indicating the standard deviation. The experimental results are plotted as purple point. The experimental data is averaged over 10 independent phase screen realisations, with the error bars indicating the standard deviation. We see excellent agreement between theory, simulation and experiment, indicating the experiment setup and the encoded phase screen are working as expected and the encoded turbulence strength closely matches the real distortions strength experienced by the system.

\section{methods}
To model a static channel under discrete, time-independent phase perturbations, each realisation $i$ is treated independently. For every realisation, a density matrix $\rho_i$ is measured and the spatially varying Bloch vectors $\mathbf{b}(\mathbf{r})$ are reconstructed. Following the coordinate-alignment and normalisation protocols detailed previously, the skyrmion number $N$ is extracted via Equation~(\ref{eq:Skyrme Number}). This procedure is repeated for each of the $n=10$ realisations per turbulence strength. To model a dynamic channel where fluctuations occur faster than the detector integration time, the system is evaluated as an ensemble-average. For each turbulence strength $\Omega$, an ensemble-averaged density matrix, $\langle \rho \rangle = \frac{1}{n}\sum_{i=1}^{n} \rho_i$, is constructed from the ten independent realisations. The spatially varying Bloch vectors are then reconstructed from this single mixed state, and the resulting skyrmion number $N$ is measured after applying the same alignment and normalisation procedures shown in Fig.~\ref{fig:No centring}.

\section{Extended data}
\subsection{Extended Dataset for Additional Topologies}
\begin{figure*}[t!]
    \includegraphics[width=0.9\textwidth]{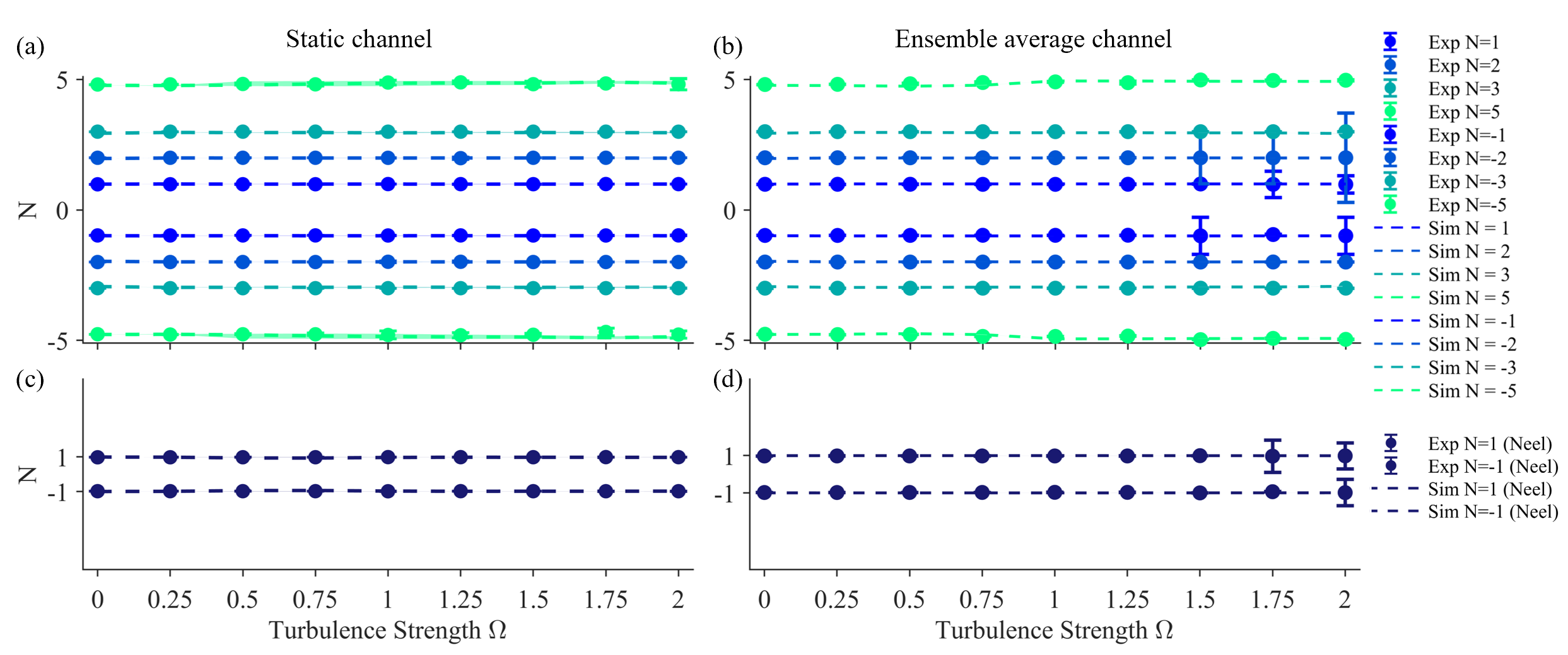}
    \caption{(a) Average measured skyrmion number $ N $ for the eight additional topologies not featured in the main text, plotted as a function of turbulence strength $\Omega$. Markers represent the mean of 10 experimental realisations, with error bars indicating the standard deviation. Dotted lines denote numerical simulations averaged over 100 realisations. (b) Ensemble-averaged skyrmion numbers $N$ extracted from the mixed state $\langle\rho\rangle$ for the same set of topologies. Each marker corresponds to the topological invariant calculated from the mean density matrix at each value of $\Omega$. Simulations (dotted lines) for the mixed-state approach show excellent convergence with the experimental data, validating the topological stability for both high-order and phase-shifted entangled states. (c)Average measured skyrmion number $N$ for the two additional phase-shifted topologies, plotted as a function of turbulence strength $\Omega$. Markers represent the mean of 10 experimental realisations, with error bars indicating the standard deviation. Dotted lines denote numerical simulations averaged over 100 realisations. (d) Ensemble-averaged skyrmion numbers $N$ extracted from the mixed state $\langle\rho\rangle$ for phase-shifted topologies. Each marker corresponds to the topological invariant calculated from the mean density matrix at each value of $\Omega$. Simulations (dotted lines) for the mixed-state approach show excellent convergence with the experimental data, validating the topological stability for both high-order and phase-shifted entangled states.}
    \label{fig:Extended data}
\end{figure*}
\noindent To demonstrate the universality of the topological protection observed in the main text, we performed a comprehensive analysis of ten distinct entangled states, spanning topological charges $N \in \{\pm 1, \pm 2, \pm 3, \pm 5\}$ as shown in Fig.~\ref{fig:Extended data}. The remaining two states are the pahse shifted states that result in a Neel type skyrmions with $N=1$ and are shown in (Fig.~\ref{fig:Extended data} (c), (d)). This extended dataset includes both pure-state realisations (Fig.~\ref{fig:Extended data} (a), (c)) and ensemble-averaged mixed states (Fig.~\ref{fig:Extended data} (b), (d)). As shown in Fig.~\ref{fig:Extended data}, the skyrmion number persists regardless of state or the specific spatial texture (e.g., Neel vs. saddle-type). The agreement between the experimental data and high-statistics numerical simulations ($n=100$) across all investigated states and turbulence strengths confirms that the robustness is a fundamental consequence of the topological mapping rather than a specific feature of a particular mode basis.

\bibliography{mybib.bib} 
\end{document}